\documentclass[aip,reprint,jap]{revtex4-1}

\usepackage{graphicx}%
\usepackage{dcolumn}%
\usepackage{bm}%

\begin{document}

\title{Promising ferrimagnetic double perovskite oxides towards high spin polarization at high temperature}

\author{Si-Da Li}
\affiliation{Beijing National Laboratory for Condensed Matter
Physics, Institute of Physics,  Chinese Academy of Sciences,
Beijing 100190, China} \affiliation{Department of Physics \&
School of Gifted Youth, University of Science and Technology of
China, Hefei 230026, China}
\author{Peng Chen}
\affiliation{Beijing National Laboratory for Condensed Matter
Physics, Institute of Physics,  Chinese Academy of Sciences,
Beijing 100190, China}
\author{Bang-Gui Liu}\email[Email:~]{bgliu@iphy.ac.cn}
\affiliation{Beijing National Laboratory for Condensed Matter
Physics, Institute of Physics,  Chinese Academy of Sciences,
Beijing 100190, China}

\date{\today}

\begin{abstract}
We predict through our first-principles calculations that four
double perovskite oxides of Bi$_2$ABO$_6$ (AB = FeMo, MnMo, MnOs,
CrOs) are half-metallic ferrimagnets. Our calculated results shows
that the four optimized structures have negative formation energy,
from -0.42 to -0.26 eV per formula unit, which implies that they
could probably be realized. In the case of Bi$_2$FeMoO$_6$, the
half-metallic gap and Curie temperature are predicted to reach to
0.71 eV and 650 K, respectively, which indicates that high spin
polarization could be kept at high temperatures far beyond room
temperature. It is believed that some of them could be synthesized
soon and would prove useful for spintronic applications.
\end{abstract}

\pacs{75.30.-m, 75.50.-y, 71.20.-b, 75.10.-b}

\maketitle

\section{Introduction}

Magnetic materials that have high spin polarization at room
temperature or higher are highly desirable for spintronic
applications\cite{wolf,cro2-96}. Half-metallic materials are good
candidates because they can have high spin polarization at high
temperature\cite{hm}. In the case of CrO$_2$, 96\% spin
polarization has been achieved experimentally\cite{cro2-96}. Since
1998, double perovskite oxides have been explored extensively for
this purpose because both half-metallicity and high Curie
temperature can be achieved in such materials\cite{lsmo,oxide}. It
has been shown experimentally that several double perovskite
oxides, such as Sr$_2$FeMoO$_6$ and Sr$_2$CrReO$_6$
materials\cite{oxide}, can keep their ferromagnetic or
ferrimagnetic phases far beyond room temperature, and more
importantly, high-quality materials have been realized recently
\cite{exp1,exp2,exp3,exp4,sum1,sum2}.

Here, we present our first-principles exploration on double
perovskite oxides of Bi$_2$ABO$_6$ with A being 3d transition
metals and B 4d/5d ones. We optimize their structures fully and
then investigate their stability, electronic structures, and
magnetic properties. We find four half-metallic ferimagnetic
materials with negative formation energies. For the best case of
Bi$_2$FeMoO$_6$, the half-metallic gap and Curie temperature reach
to 0.71 eV and 650 K, respectively. This means that high spin
polarization could be realized at high temperatures well above
room temperature. More detailed results will be presented in the
following.

\section{Computational details}

We use the pseudo-potential and plane wave methods within the
density functional theory (DFT) \cite{dft}, as implemented in
package VASP\cite{vasp}. We use generalized gradient approximation
(GGA)\cite{pbe96} for the exchange-correlation potential. In
addition to usual valence states, the semicore d states are
considered for Bi and the semicore p states for Cr, Mn, Fe, Mo,
and Os. Scalar approximation is used for relativistic
effect\cite{relsa}, and the spin-orbit coupling is neglected
because it has little effect on our main conclusion (to be
detailed in the following). We use Monkhorst-Pack method to
generates the K-point mesh\cite{mpmesh}, choosing
6$\times$6$\times$6 (6$\times$6$\times$4) for structure
optimizations and total energy calculations of 10-atom (20-atom)
unit cells and 12$\times$12$\times$12 for electronic structure
calculations. The cut-off energy is set to 500 eV and the criteria
for convergence is $10^{-6}$ eV for electronic steps and 0.005
eV/\AA{} on atoms for ionic steps.

Metropolis algorithm and variants are used for our Monte Carlo
simulations\cite{mc,metropolis}. Phase transition temperatures
$T_c$ are determined through investigating the average
magnetization, magnetic susceptibility, and fourth-order Binder's
cumulant as functions of temperature\cite{mc}. Several
three-dimensional lattices of upto 30$\times$30$\times$30 magnetic
unit cells with periodic boundary condition are used in these
calculations. The first 90,000 Monte Carlo steps (MCS) of total
150,000 MCS are used for the thermal equilibrium, and the
remaining 60,000 MCS are used to calculate the average
magnetization for a given temperature. The Curie temperature $T_c$
value is determined through investigating the magnetization as a
function of temperature.

\section{Main calculated results and analysis}

Comparing Sr$_2$FeMoO$_6$ \cite{oxide}, Bi$_2$FeCrO$_6$
\cite{exp1,exp2,bi2fecro6,bi2fecro6a} and others similar
\cite{exp3,exp4,sum1,sum2}, we consider double perovskite
structure of formula Bi$_2$ABO$_6$, taking some 3d
transition-metal elements for A and some 4d/5d for B. We optimize
fully their crystal structures in terms of the unit cell of the 10
atoms. The optimized Bi$_2$ABO$_6$ has space group Rc (\#146).
This crystal structure, similar to R3c (\#161), is distorted from
cubic double perovskite structure, has rhombohedral symmetry, and
includes 10 internal parameters\cite{oxide,sum1,sum2}. We shall
present four of the Bi$_2$ABO$_6$ compounds, for AB = FeMo, MnMo,
MnOs, and CrOs, because they have negative formation energies so
that their experimental realization should be probable. Our
optimized structural parameters of the four Bi$_2$ABO$_6$
compounds are summarized in Table I. The total magnetic moments
per formula unit and the partial moments in the spheres of the
magnetic A and B atoms are summarized in Table II. The magnetic
moments in the spheres of other atoms are much smaller. The total
moments are integers in unit of Bohr magneton $\mu_B$, showing a
feature of half-metallicity\cite{hm}. The magnetic moment at the A
atom is antiparallel to that at the B atom, which means that the
magnetic order in these compounds is ferrimagnetic.

\begin{table}[htb]
\caption{Optimized structural parameters of double perovskite
Bi$_2$ABO$_6$ with the Rc (\#146) crystal structure for AB = FeMo,
MnMo, MnOs, and CrOs.}\label{table1}
\begin{ruledtabular}
\begin{tabular}{lcccc}
AB & FeMo & MnMo & MnOs &  CrOs  \\
\hline
 $a$ (\AA) & 5.725  & 5.779  & 5.761 & 5.771   \\
 $c$ (\AA) & 14.054 & 14.093 & 13.607 & 12.816  \\
$\alpha$ ($^\circ$)  &  59.91 & 60.20  & 61.61  & 64.37 \\
\hline
Bi $z_1$  & 0.9881 & 0.9859  &  0.9917 & 0.0294   \\
Bi $z_1^\prime$ & 0.4837 & 0.4851  & 0.4923  & 0.5071 \\
A  $z_2$   & 0.2589 & 0.2572  & 0.2593  & 0.2682   \\
B  $z_2^\prime$ & 0.7675 & 0.7657  & 0.7647  & 0.2681   \\
O $x_3$    & 0.5504 & 0.5612  & 0.5534  & 0.4811   \\
O $x_3^\prime$ & 0.0506 & 0.0471  & 0.0520  & 0.0558   \\
O $y_3$    & 0.9357 & 0.9220  & 0.9273  & 0.9353  \\
O $y_3^\prime$ & 0.4350 & 0.4397  & 0.4348  & 0.4287  \\
O $z_3$    & 0.1022 & 0.1036  & 0.0981  & 0.1075  \\
O $z_3^\prime$ & 0.6099 & 0.6145  & 0.6085  & 0.6009  \\
 \end{tabular}
 \end{ruledtabular}
\end{table}

In Fig. 1 we present spin-resolved density of states (DOS, in
states/eV per formula unit) between -7.7 and 3 eV of the double
perovskite Bi$_2$FeMoO$_6$ and Bi$_2$MnMoO$_6$. The total DOS in
majority-spin channel is zero at the Fermi level in both of the
cases. This indicates that the two double perovskite compounds are
half-metallic, in agreement with the integral magnetic moments in
unit of $\mu_B$. In Fig. 2 we present spin-resolved density of
states between -8 and 3 eV of the double perovskite
Bi$_2$MnOsO$_6$ and Bi$_2$CrOsO$_6$. They are both half-metallic,
too, but it is in minority-spin channel that the total DOS at the
Fermi level is equivalent to zero in these two cases. The filled
electronic states near the Fermi level originate mainly from the B
atom (Mo or Os). We can use half-metallic gap $E_g$ as the key
parameter to describe the half-metallic
property\cite{hm,lbg1,lbg2,lbgbook}. The $E_g$ values of the four
compounds, from 0.25 to 0.71 eV, are summarized in Table II. For
the Bi$_2$FeMoO$_6$, $E_g$ is equivalent to 0.71 eV, which implies
that high spin polarization could be robust even after the
spin-orbit coupling is taken into account.

\begin{figure}[!htb]
\includegraphics[width=7cm]{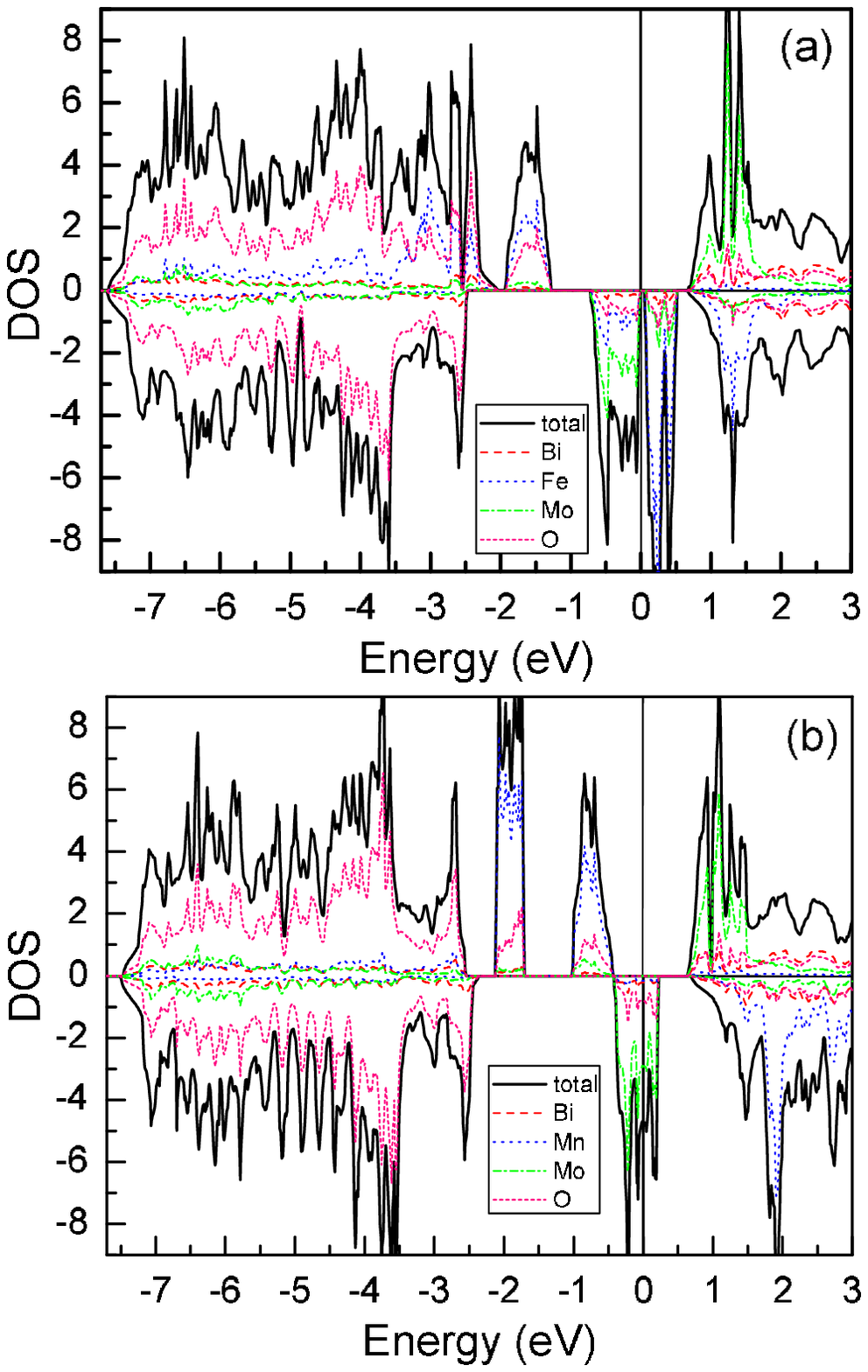}
\caption{(color online) Spin-resolved density of states (DOS, in
state/eV per formula unit) of double perovskite Bi$_2$ABO$_6$ for
AB=FeMo (a) and AB=MnMo (b). The solid line is total DOS, and
short-dashed, dot-dashed, and dotted lines refer to partial DOS
projected in the atomic spheres of Bi, A, B, and O, respectively.
The upper part in each panel is majority-spin DOS result, and the
lower the minority-spin one.}\label{dos1}
\end{figure}

\begin{figure}[!htb]
\includegraphics[width=7cm]{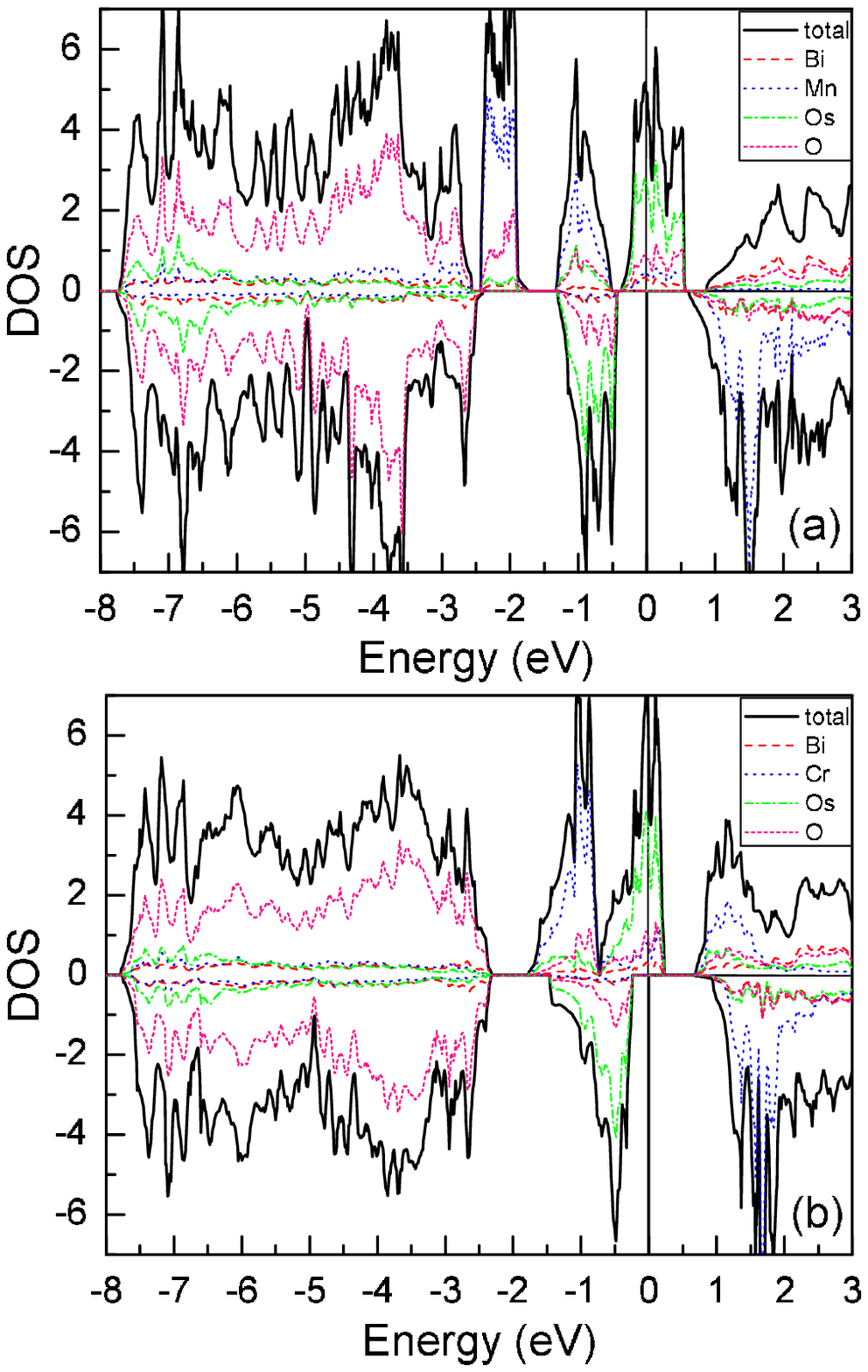}
\caption{(color online) Spin-resolved density of states (DOS, in
state/eV per formula unit) of double perovskite Bi$_2$ABO$_6$ for
AB=MnOs (a) and AB=CrOs (b). The solid line is total DOS, and
short-dashed, dot-dashed, and dotted lines refer to partial DOS
projected in the atomic spheres of Bi, A, B, and O, respectively.
The upper part in each panel is majority-spin DOS result, and the
lower the minority-spin one.}\label{dos2}
\end{figure}

We investigate their formation energies to determine the stability
of these materials towards experimental realization. For achieving
reasonable reliability, we choose stable and reachable compounds
as our references, and try to use those reference compounds whose
valence states concerned are close to those of our compounds.
Therefore, we use $\mathrm{Bi_2O_3}$, $\mathrm{Cr_3O_4}$,
$\mathrm{Mn_3O_4}$, $\mathrm{Fe_3O_4}$, $\mathrm{MoO_2}$, and
$\mathrm{OsO_2}$ for our reference compounds for calculating the
formation energies. The formation energy is defined as
\begin{equation}
E_f = E(\mathrm{Bi_2 ABO_6})-E_{\mathrm{ref}},
\end{equation}
where $E(\mathrm{X})$ is the total energy of X, and
$E_{\mathrm{ref}}$ is defined as $E(\mathrm{Bi_2O_3 })+\frac 13
E(\mathrm{A_3O_4 })+2E(\mathrm{BO_2})-\frac 76 E(\mathrm{O_2})$.
This criteria is much more severe than merely using AO compound or
bulk A materials because O atom in the gas state has higher energy
than in compounds such as $\mathrm{Fe_3O_4}$. This should be more
precise because the bonds in our materials are almost formed
between metal atom and O, not between metal atoms. The formation
energies for the four compounds are summarized in Table II. The
negative values means that they should probably be realized.

\begin{table}[!htb]
\caption{Calculated values of formation energy ($E_f$ in eV),
magnetic moment of atom A ($M_A$ in $\mu_B$), magnetic moment of
atom B ($M_B$ in $\mu_B$), total magnetic moment ($M$ in $\mu_B$
per formula unit), half-metallic gap ($E_g$ in eV), and Curie
temperature ($T_c$ in K) of double perovskite Bi$_2$ABO$_6$ for AB
= FeMo, MnMo, MnOs, and CrOs.}\label{table1}
\begin{ruledtabular}
\begin{tabular}{ccccc}
AB & FeMo & MnMo & MnOs &  CrOs  \\
\hline
$M_A$ & 3.638 & 4.279 & 4.066 & 2.636 \\
$M_B$ & -1.755& -1.391& -1.041& -0.613 \\
$M$   & 2.000 & 3.000 & 3.000 & 2.000 \\
\hline
$E_g$ & 0.71  & 0.47  & 0.46  & 0.25 \\
\hline
$E_f$ & -0.41 & -0.26 & -0.42 & -0.29 \\
\hline
$T_c$ & 650 & 255 & 174 & 201 \\
 \end{tabular}
 \end{ruledtabular}
\end{table}

In order to estimate the Curie temperatures ($T_c$) of the
materials, we calculate the spin exchange interactions between the
nearest and next nearest neighboring magnetic atoms (A and B) in
terms of the 20-atom unit cells. Rigorously speaking, there are
some induced spin density in the spheres of the Bi and O atoms,
less than $0.05\mu_B$. Because they are very small compared to
those in the spheres of the magnetic atoms, we shall consider only
the magnetic atoms in the following. Actually, A and B atoms form
a lattice of magnetic unit cells (cubic unit cells of a NaCl
crystal structure) \cite{oxide,sum1,sum2}. In these calculations,
we fix the structures and change the magnetic orders of A and B
atoms. In order to make the electronic steps converge for a
magnetic order, the linear mixing parameter should be decreased to
an small value, 0.1 or smaller. Through comparing the total
energies, we obtain the spin exchange interaction constants:
$J_{\mathrm{AB}}$ for the nearest A and B pair, $J_{\mathrm{AA}}$
and $J_{\mathrm{BB}}$ for the A-A and B-B next nearest pairs.
$J_{\mathrm{AB}}$ is dominant over the others. The resultant spin
Hamiltonian reads:
\begin{equation}
H=\sum_{\langle ij\rangle}J_{ij}\vec{S}_i\cdot \vec{S}_j
\end{equation}
where $\vec{S}_i$ is spin operator at site $i$ (in both of the A
and B sublattices), the summation is over spin pairs, and the spin
interaction constant $J_{ij}$ is limited to the nearest and the
next nearest neighboring spins.

\begin{figure}[!htb]
\includegraphics[width=7cm]{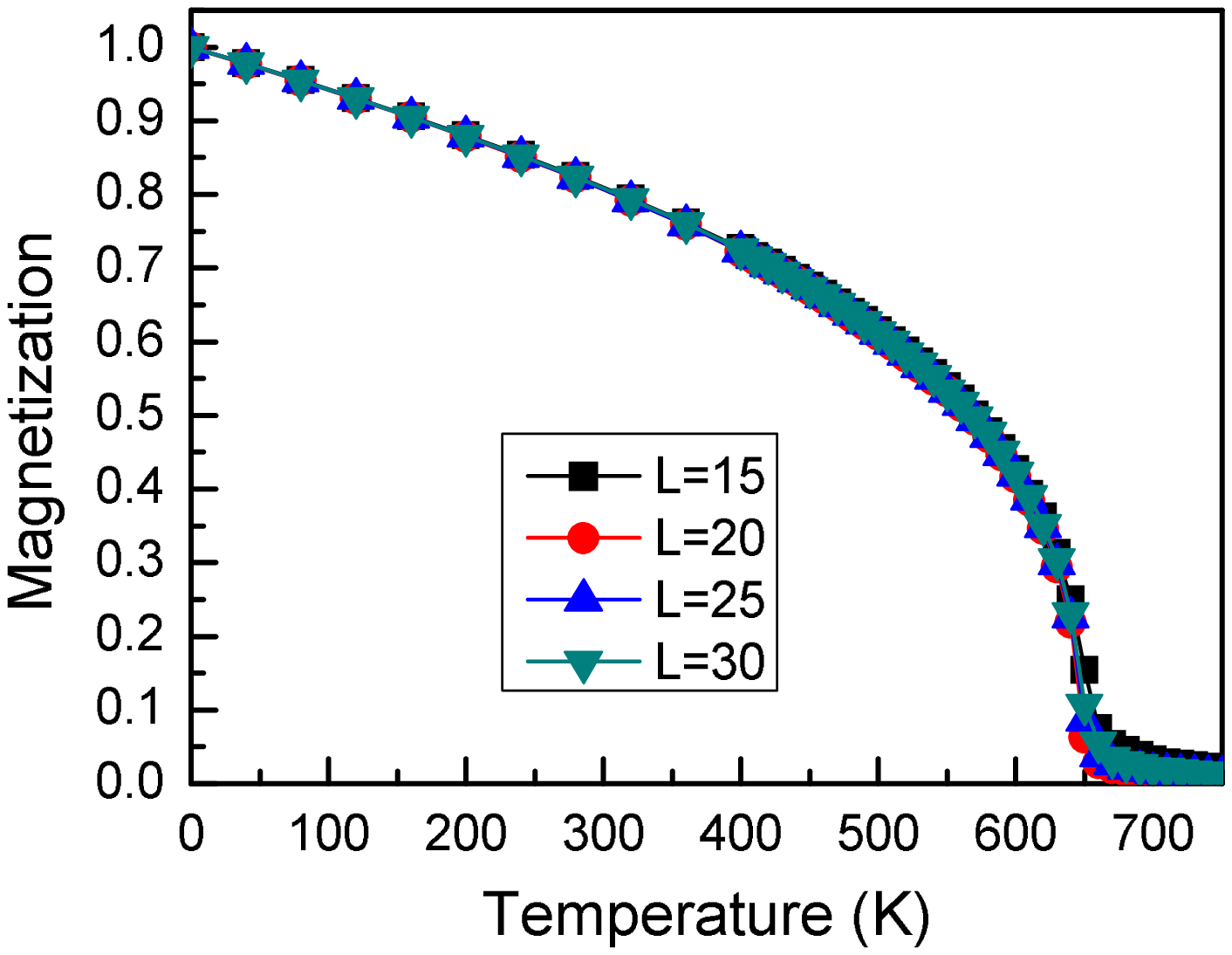}
\caption{(color online) Average normalized magnetizations as
functions of temperature for double perovskite Bi$_2$FeMoO$_6$ for
four different $L$ values. Monte Carlo simulations are done with
$L\times L\times L$ magnetic unit cells.}\label{mag}
\end{figure}

We carry out Monte Carlo simulations to estimate the $T_c$ of the
materials\cite{mc,metropolis}. It is well known that Curie
temperature will be a little underestimated if classical
approximation to the Heisenberg model (2) is used in the Monte
Carlo simulation, but it must be much overestimated if the model
(2) is reduced to Ising model. For comparison, we do our Monte
carlo simulations with both of the approximate models. We present
the average normalized magnetization of Bi$_2$FeMoO$_6$, from
classical Heisenberg model, as a representative in Fig. 3. The
$T_c$ value can be estimated to be 650 K. The others can be done
in the same way. The calculated $T_c$ values are summarized in
Table II. In contrast, the Ising model results are 1010, 396, 264,
and 270 K for AB = FeMo, MnMo, MnOs, and CrOs, respectively.
Therefore, the Curie temperatures for the four half-metallic
ferrimagnets are, at least, 650, 255, 174, and 201 K for
Bi$_2$FeMoO$_6$, Bi$_2$MnMoO$_6$, Bi$_2$MnOsO$_6$, and
Bi$_2$CrOsO$_6$, respectively. High Curie temperature well above
room temperature could be realized in Bi$_2$FeMoO$_6$.

\section{Discussions and Conclusion}

Our calculated results shows that the spin exchange interaction
between the nearest A and B atoms is positive, and the A-A and B-B
interactions are either weak or negative depending on specific A
and B atoms\cite{sum1,sum2}. In the case of Bi$_2$FeMoO$_6$, our
calculations show that the nearest A-B spin exchange energy is
39.2 meV, and the nearest A-A and B-B spin exchange energies are
0.13 and -0.71 meV, respectively. The main spin interaction is
intermediated by the O atom in between the magnetic A and B atoms,
with the A-O-B bond angle being almost 180$^{\circ}$, and
therefore, it is an antiferromagnetic superexchange. The A atom
contributes a different magnetic moment from the B atom so that
ferrimagnetism is formed in these double perovskite compounds.
Possible overlapping of the nearest O wave functions should play
some roles in these compounds, but the main mechanism for the
ferrimagnetism must be the antiferromagnetic superexchange between
the nearest A and B atoms.

In summary, our first-principles calculations show that four
double perovskite oxides, Bi$_2$ABO$_6$ (AB = FeMo, MnMo, MnOs,
and CrOs), have negative formation energy, from -0.42 to -0.26 eV
per formula unit. In the case of Bi$_2$FeMoO$_6$, our calculated
results uncover that its half-metallic gap and Curie temperature
reach to 0.71 eV and 650 K, respectively. These indicates that
they could probably be realized and high spin polarization could
be achieved at high temperature. We believe that at least some of
them could be synthesized soon and would prove useful for
spintronic applications.

\begin{acknowledgments}
This work is supported by Chinese Department of Science and
Technology (Grant No. 2012CB932302) and by Nature Science
Foundation of China (Grant Nos. 11174359 and 10874232).
\end{acknowledgments}

\end{document}